\documentclass[12pt,a4paper]{article}
\pdfoutput=1
\usepackage{color}
\usepackage{amssymb,amsmath,bm,bbold}
\usepackage{epsf}
\usepackage{epsfig}
\usepackage{relsize}
\usepackage[dvipsnames]{xcolor}
\usepackage[linktoc=page,bookmarks=false,colorlinks=false,linkbordercolor=RoyalBlue,citebordercolor=ForestGreen,urlbordercolor=CornflowerBlue]{hyperref}
\usepackage{latexsym,mathrsfs,dsfont}
\usepackage[normalem]{ulem} 
\usepackage[compress]{cite}
\usepackage{graphicx}
\usepackage{url}
\usepackage{booktabs}
\usepackage{float}
\usepackage{multirow}
\usepackage{changepage}
\usepackage[hypcap]{caption, subcaption}

\usepackage[titles]{tocloft}

\usepackage{tabularx,colortbl}

\usepackage{fancyhdr}
\advance \headheight by 3.0truept       

\allowdisplaybreaks[1]

\definecolor{red}{cmyk}{0,1,1,0.4}
\definecolor{darkgreen}{rgb}{0.0,0.6,0.0}
\definecolor{cDarkGrey}{RGB}{91,91,91}
\definecolor{cGrey}{RGB}{245,243,238}
\definecolor{cBlue}{RGB}{0,110,191}
\definecolor{cLightBlue}{RGB}{214,237,252}
\definecolor{cRed}{RGB}{196,0,100}
\definecolor{cLightRed}{RGB}{254,222,237}
\definecolor{cGreen}{RGB}{0,166,80}
\definecolor{cLightGreen}{RGB}{254,222,237}
\definecolor{cOrange}{RGB}{221,74,44}
\definecolor{cLightOrange}{RGB}{255,215,210}
\definecolor{cPurple}{RGB}{93,35,125}
\definecolor{cLightPurple}{RGB}{241,230,252}
\definecolor{cYellow}{RGB}{252,191,10}
\definecolor{cISSRBlue}{RGB}{0,111,174}
\definecolor{cISSRGrey}{RGB}{167,169,172}

\newcommand{\beq}{\begin{equation}}
\newcommand{\eeq}{\end{equation}}
\newcommand{\be}{\begin{equation}}
\newcommand{\ee}{\end{equation}}
\newcommand{\bi}{\begin{itemize}}
\newcommand{\ei}{\end{itemize}}
\newcommand{\ba}{\begin{array}}
\newcommand{\ea}{\end{array}}
\newcommand{\beqa}{\begin{eqnarray}}
\newcommand{\eeqa}{\end{eqnarray}}
\newcommand{\bea}{\begin{eqnarray}}
\newcommand{\eea}{\end{eqnarray}}
\newcommand{\beqn}{\begin{eqnarray}}
\newcommand{\eeqn}{\end{eqnarray}}

\newcounter{TODO}

\newcommand{\vcb}{|V_{cb}|}

\newcommand{\vub}{|V_{ub}|}

\newcommand{\vus}{|V_{us}|}

\def\kpn{K^+\rightarrow\pi^+\nu\bar\nu}

\begin{document}

\begin{flushleft}
\end{flushleft}

\vspace{-14mm}
\begin{flushright}
  AJB-23-5
\end{flushright}

\medskip

\begin{center}
{\Large\bf\boldmath
  Waiting for Precise Measurements of $\beta$ and $\gamma$  
}
\\[1.0cm]
{\bf\large
    Andrzej~J.~Buras
}\\[0.3cm]

{\small
TUM Institute for Advanced Study,
    Lichtenbergstr. 2a, D-85748 Garching, Germany \\[0.2cm]
Physik Department, TUM School of Natural Sciences, TU M\"unchen,\\ James-Franck-Stra{\ss}e, D-85748 Garching, Germany
}
\end{center}

\vskip 0.5cm

\begin{abstract}
  \noindent
  During the last three decades the determination of the Unitarity Triangle (UT)
  was dominated by the measurements of its sides $R_b$ and $R_t$ through
  tree-level $B$ decays and the $\Delta M_d/\Delta M_s$ ratio, respectively,
  with
  some participation of the measurements of the angle $\beta$ through the mixing
  induced   CP-asymmetries like $S_{\psi K_S}$ and $\varepsilon_K$. However, as pointed out already in
  2002 by Fabrizio Parodi, Achille Stocchi and the present author, the most
  efficient strategy for a precise determination of the apex of the UT, that is
  $(\bar\varrho,\bar\eta)$, is to use the measurements of the angles $\beta$ and $\gamma$. The second best strategy would be the measurements of $R_b$ and $\gamma$. However, in view of the tensions between different determinations of $|V_{ub}|$ and  $|V_{cb}|$, that enter $R_b$,  the $(\beta,\gamma)$ strategy should be   a clear winner once LHCb and Belle II will improve  the measurements of these two angles.   In this note we recall our finding of 2002 which should be finally realized
  in this decade through precise measurements of both angles by these collaborations. In this context we present two very simple formulae
  for $\bar\varrho$ and $\bar\eta$ in terms of $\beta$ and $\gamma$ which could be
  derived by high-school students, but to my  knowledge never appeared in
  the literature on the UT, not even in our 2002 paper. We also emphasize the
  importance of precise measurements of both angles that would allow to
  perform powerful tests of the SM through numerous  $|V_{cb}|$-independent correlations   between $K$ and $B$ decay branching ratios $R_i(\beta,\gamma)$ recently
  derived by Elena Venturini and the present author. The simple findings presented here
  will appear in a subsection of a much longer  contribution to the proceedings
  of KM50 later this year. I exhibited them here so that they are not lost in  the latter.
  
  \end{abstract}

\thispagestyle{empty}
\newpage
\setcounter{page}{1}
\section{Introduction}
\begin{figure}
\centering
\includegraphics[width = 0.55\textwidth]{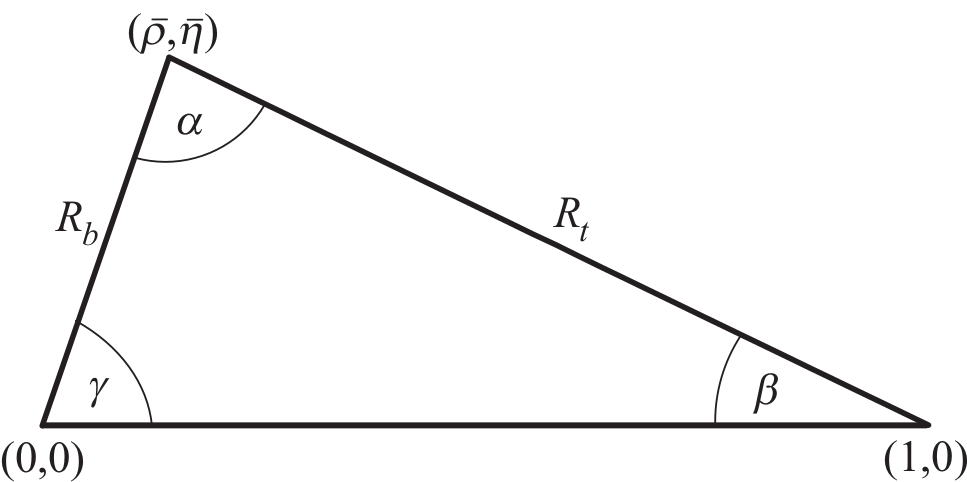}
 \caption{\it The Unitarity Triangle. }\label{UUTa}
\end{figure}

This year we celebrate the 60th anniversary of the Cabibbo angle \cite{Cabibbo:1963yz} and the 50th aniversary of the Kobayashi-Maskawa paper  \cite{Kobayashi:1973fv} in which they predicted the existence of the third generation of quarks,
required for the explanation of the observed CP-violation within the Standard Model (SM). In this seminal paper they also presented a parametrization of  a unitary $3\times 3$ matrix that quaranteed the absence of FCNC processes at the tree-level in the SM through Glashow, Iliopoulos and
Maiani (GIM) mechanism \cite{Glashow:1970gm}. There are different parametrizations of this matrix, in particular the standard parametrization of the CKM matrix of Chau and Keung \cite{Chau:1984fp}, the approximate  Wolfenstein parametrization \cite{Wolfenstein:1983yz} and  much more precise Wolfenstein-like parametrization introduced in Munich in 1994 \cite{Buras:1994ec} in which in particular the apex of the unitarity triangle (UT) in Fig.~\ref{UUTa} is described by $(\bar\varrho,\bar\eta)$. The latter parametrization dominates present CKM phenomenology.

All these parametrizations involve three real parameters and one complex phase.
In particular in 1986 Harari and Leurer \cite{Harari:1986xf} recommended
the standard parametrization because of the relations\footnote{$c_{13}=1$ to an excellent accuracy.}
\be\label{LH}
s_{12}=\vus, \qquad   s_{13}=\vub,\qquad s_{23}=\vcb, \qquad \gamma,
\ee
 that allows the determination of  these  four parameters separately in tree-level decays.
Consequently,  basically all flavour phenomenology in the last three decades  used such sets of parameters. In particular the determination of the UT
  was dominated by the measurements of its sides $R_b$ and $R_t$ through
  tree-level $B$ decays and the $\Delta M_d/\Delta M_s$ ratio, respectively, with   some participation of the measurements of the angle $\beta$ through the mixing  induced CP-asymmetries like $S_{\psi K_S}$, the parameter $\varepsilon_K$  and much less precise angle $\gamma$. This is the case not only
  of global analyses by UTfitter\cite{Bona:2007vi}, CKMfitter \cite{Charles:2004jd}  and PDG \cite{Workman:2022ynf} but also of less sophisticated
  determinations of the CKM matrix and of the UT.

  However, as pointed out in  \cite{Blanke:2018cya,Buras:2022wpw,Buras:2021nns,Buras:2022qip}, the most powerful strategy  appears eventually to be the one which
  uses 
as basic CKM parameters
\be\label{BBV}
\vus=\lambda, \qquad \vcb,\qquad  \beta,\qquad \gamma,
\ee
that is two mixing angles and two phases with $\beta$ replacing $\vub$ in
(\ref{LH}). In my view this choice is superior to the one in (\ref{LH}) for
several reasons:
\begin{itemize}
\item
  The known tensions between exclusive and inclusive determinations of $\vcb$
  and $\vub$ \cite{Bordone:2021oof,FlavourLatticeAveragingGroupFLAG:2021npn}
  are represented only by $\vcb$ which can be eliminated efficiently
  by constructing suitable ratios of flavour observables $R_i(\beta,\gamma)$
  which are free of the tensions in question. These ratios 
  are often independent of these two phases or dependent only on one of them \cite{Buras:2022wpw,Buras:2021nns,Buras:2022qip}. See Table\,4 in \cite{Buras:2021nns}.
\item
  As pointed out already in 2002 \cite{Buras:2002yj},
  the most
  efficient strategy for a precise determination of the apex of the UT, that is
  $(\bar\varrho,\bar\eta)$, is to use the measurements of the angles $\beta$ and $\gamma$. Indeed, among any pairs of two variables representing the sides and the angles of the UT that are assumed for this exercise to be known with the same precision, the measurement of $(\beta,\gamma)$ results in the most accurate values of $(\bar\varrho,\bar\eta)$. The second best strategy would be the measurements of $R_b$ and $\gamma$. However, in view of the tensions between different determinations of $\vub$ and  $\vcb$, that enter $R_b$,  the $(\beta,\gamma)$ strategy will  be a clear winner once LHCb and Belle II collaborations will improve  the measurements of these two angles.
\item
  The $\vcb-\gamma$ plots for fixed $\beta$ proposed in  \cite{Buras:2022wpw,Buras:2021nns} are, as emphasized in \cite{Buras:2022nfn}, useful companions to
  common UT fits because they exhibit better possible inconsistencies between $\vcb$ and $(\beta,\gamma)$ determinations than the latter fits.
\end{itemize}

In this note we would like to emphasize the importance
of the $(\beta,\gamma)$ strategy and  present new formulae for $\bar\varrho$ and
$\bar\eta$ in terms of these two UT parameters that are accompanied
by a simple numerics. Finally we recall briefly the prospects for precise measurements
of $\beta$ and $\gamma$ in the near future.

\boldmath
\section{New Formulae for $\bar\varrho$ and $\bar\eta$}
\unboldmath
We begin with two simple formulae that are central in the
$(\beta,\gamma)$ strategy but 
to my knowledge have not been presented in the literature before,
not even in our 2002 paper \cite{Buras:2002yj}. They read
\be\label{AJB23}
\boxed{\bar\varrho=\frac{\sin\beta\cos\gamma}{\sin(\beta+\gamma)},\qquad
  \bar\eta=\frac{\sin\beta\sin\gamma}{\sin(\beta+\gamma)}}
\ee
so that $(\bar\varrho,\bar\eta)$ can be be found in no time once
 $\beta$ and $\gamma$ are known.
They  follow simply from
\be
\bar\varrho=R_b\cos\gamma,\qquad \bar\eta=R_b\sin\gamma, \qquad
R_b=\frac{\sin\beta}{\sin(\beta+\gamma)}
\ee
with the first two relations representing $(R_b,\gamma)$ strategy \cite{Buras:2002yj}. Evidently these formulae can be derived by high-school students\footnote{In fact I recall that I had to solve such a triangle problem in 1962.}, but the UT is unknown to them and somehow to my knowledge no flavour physicist got the idea to present them in print so far.

While precise measurements of $\beta$ and $\gamma$ are important for the determination of the UT, they will also be very important for the tests of strategies
developed in \cite{Buras:2022wpw,Buras:2021nns,Buras:2022qip}.
Indeed as demonstrated recently in these papers for $K$ decays
and in the case of $B$ decays already in 2003  \cite{Buras:2003td}
one can construct a multitude of $\vcb$-independent ratios $R_i(\beta,\gamma)$ not only
of decay branching ratios to quark mixing observables but also of branching
ratios themselves. Those which involve branching ratios from different meson
systems depend generally on $\beta$ and $\gamma$. Once $\beta$ and $\gamma$
will be precisely measured, this multitude of the ratios $R_i(\beta,\gamma)$ will provide very good tests of the SM independently of the uncertain values of $\vcb$ and $\vub$ from tree-level decays. Judging from the development in the
last ten years, it could still take some time before they are reduced to
a satisfactory level.

The explicit expressions for $R_i(\beta,\gamma)$ can be found in 
in \cite{Buras:2022wpw,Buras:2021nns,Buras:2022qip} together with plots
for them as functions of $\beta$ and $\gamma$. Moreover, determining CKM parameters from $\Delta F=2$ process only, thereby avoiding the $\vcb$ tensions,
all ratios $R_i(\beta,\gamma)$ can be predicted in the SM rather precisely already today. We quote here only  \cite{Buras:2022qip}
 \be\label{R1}
\boxed{\frac{\mathcal{B}(\kpn)}{\left[{\overline{\mathcal{B}}}(B_s\to\mu^+\mu^-)\right]^{1.4}}= 53.69\pm2.75\,,}
\ee
\be\label{R5}
\boxed{\frac{\mathcal{B}(\kpn)}{\left[\mathcal{B}(B^+\to K^+\nu\bar\nu)\right]^{1.4}}=(1.90\pm0.13)\times 10^{-3}\,}
\ee
that could be of particular interest  when the measurements of $\kpn$, $B_s\to\mu^+\mu^-$ and $B^+\to K^+\nu\bar\nu$ will be improved by NA62, CMS, LHCb and ATLAS
at CERN and Belle II experiment at KEK. Results for 26 $K$ and $B$ decay branching ratios using
this strategy are listed in \cite{Buras:2022wpw,Buras:2022qip}.

The corresponding CKM parameters are found to be \cite{Buras:2022wpw}
\be\label{CKMoutput}
\boxed{\vcb=42.6(4)\times 10^{-3}, \qquad 
\gamma=64.6(16)^\circ, \qquad \beta=22.2(7)^\circ,}
\ee
and consequently
\be\label{CKMoutput3}
\boxed{\bar\varrho=0.164(12),\qquad \bar\eta=0.341(11)\,.}
\ee

In Tables~\ref{tab:1} and \ref{tab:2} we show $\bar\varrho$ and $\bar\eta$ as functions on $\gamma$ for different values of $\beta$ in the expected ranges
for the latter parameters. The present experimental determinations of $\beta$
and $\gamma$ read
\be\label{betagamma}
\boxed{\beta=(22.2\pm 0.7)^\circ, \qquad  \gamma = (63.8^{+3.5}_{-3.7})^\circ \,,} \ee
with the first one used in (\ref{CKMoutput3}).  Here the value for $\gamma$ is the most recent one from the LHCb which updates
the one in \cite{LHCb:2021dcr} $(65.4^{+3.8}_{-4.2})^\circ$. It is not as precise
as the one in (\ref{CKMoutput}) but fully consistent with it. In the coming
years LHCb and Belle II should reduce the error in $\gamma$ in the ballpark
of $1^\circ$ and also the error on $\beta$ will be reduced. In fact for the
strategies of  \cite{Buras:2022wpw,Buras:2021nns,Buras:2022qip} the reduction
of the error on $\beta$ is even more important than the one on $\gamma$ because
$\beta$ is an input but $\gamma$ an output.

It is interesting to compare these tables with the most recent ``angle-only fit'' of UTfitter \cite{UTfit:2022hsi}
\be\label{CKMoutput5}
{\bar\varrho=0.156(17),\qquad \bar\eta=0.341(12)\,.}
\ee
The significant error on $\bar\varrho$ will be reduced by much with the improved measurement of $\gamma$. Let us then anticipate a future measurements with
\be\label{CKMoutput26}
\gamma=64.0(10)^\circ, \qquad \beta=22.2(4)^\circ,\qquad (2026)\,.
\ee
Using (\ref{AJB23}) and adding the errors in quadrature we find
\be
\bar\varrho=0.166(7),\qquad \bar\eta=0.340(6)\,,\qquad (2026)\,.
\ee
Once the tensions in the determination of $\vcb$ and $\vub$ will be resolved
the full fit will of course result in even more precise values.

\begin{table}[!ht]
    \centering
    \begin{tabular}{|l|l|l|l|l|l|l|l|l|}
    \hline
        $\gamma/\beta$ & $21.6^\circ$ & $21.8^\circ$ & $22.0^\circ$ & $22.2^\circ$ &
$22.4^\circ$ & $22.6^\circ$ &$22.8^\circ$ & $23.0^\circ$\\ \hline
        $ 60^\circ$ & 0.186 & 0.188 & 0.189 & 0.191 & 0.192 & 0.194 & 0.195 & 0.197 \\
        $ 61^\circ$ & 0.180 & 0.181 & 0.183 & 0.184 & 0.186 & 0.187 & 0.189 & 0.190 \\
        $ 62^\circ$ & 0.174 & 0.175 & 0.177 & 0.178 & 0.180 & 0.181 & 0.183 & 0.184 \\
        $ 63^\circ$ & 0.168 & 0.169 & 0.171 & 0.172 & 0.174 & 0.175 & 0.176 & 0.178 \\
        $ 64^\circ$ & 0.162 & 0.163 & 0.165 & 0.166 & 0.167 & 0.169 & 0.170 & 0.172 \\
        $ 65^\circ$ & 0.156 & 0.157 & 0.159 & 0.160 & 0.161 & 0.163 & 0.164 & 0.165 \\
        $ 66^\circ$ & 0.150 & 0.151 & 0.152 & 0.154 & 0.155 & 0.156 & 0.158 & 0.159 \\
        $ 67^\circ$ & 0.144 & 0.145 & 0.146 & 0.148 & 0.149 & 0.150 & 0.151 & 0.153 \\
        $ 68^\circ$ & 0.138 & 0.139 & 0.140 & 0.142 & 0.143 & 0.144 & 0.145 & 0.146 \\ \hline
    \end{tabular}
    \caption{\it Values of $\bar\varrho$ for different values of $\beta$ and $\gamma$ .}
    \label{tab:1}
\end{table}

\begin{table}[!ht]
    \centering
    \begin{tabular}{|l|l|l|l|l|l|l|l|l|}
    \hline
       $\gamma/\beta$ & $21.6^\circ$ & $21.8^\circ$ & $22.0^\circ$ & $22.2^\circ$ &
$22.4^\circ$ & $22.6^\circ$ &$22.8^\circ$ & $23.0^\circ$\\
        \hline
         $ 60^\circ$ &$ 0.322$ &$ 0.325$ &$ 0.328 $&$ 0.330 $&$ 0.333 $&$ 0.336 $&$ 0.338$ &$ 0.340$\\
         $ 61^\circ$ & 0.325 & 0.327 & 0.330 & 0.333 & 0.336 & 0.338 & 0.341 & 0.344 \\
         $ 62^\circ$ & 0.327 & 0.330 & 0.333 & 0.335 & 0.338 & 0.341 & 0.344 & 0.346 \\
         $ 63^\circ$ & 0.329 & 0.332 & 0.335 & 0.338 & 0.341 & 0.343 & 0.346 & 0.349 \\
         $ 64^\circ$ & 0.332 & 0.335 & 0.338 & 0.340 & 0.343 & 0.346 & 0.349 & 0.352 \\
         $ 65^\circ$ & 0.334 & 0.337 & 0.340 & 0.343 & 0.346 & 0.349 & 0.351 & 0.354 \\
         $ 66^\circ$ & 0.337 & 0.340 & 0.342 & 0.345 & 0.348 & 0.351 & 0.354 & 0.357 \\
         $ 67^\circ$ & 0.339 & 0.342 & 0.345 & 0.348 & 0.351 & 0.354 & 0.357 & 0.360 \\
         $ 68^\circ$ & 0.341 & 0.344 & 0.347 & 0.350 & 0.353 & 0.356 & 0.359 & 0.362 \\ \hline
    \end{tabular}
    \caption{\it Values of $\bar\eta$ for different values of $\beta$ and $\gamma$ .}
    \label{tab:2}
\end{table}

\section{Outlook}
During this decade significant progress in the measurements of $\beta$ and in particular of $\gamma$ is expected. A measurement of $\gamma$ with an accuracy of $1^\circ$ should be possible. Various strategies for achieving this goal, in particular using $B\to DK$ decays, are reviewed in the  Belle II book \cite{Belle-II:2018jsg}, in chapter 8 of
my book \cite{Buras:2020xsm} and in particular in the classic paper by Robert Fleischer \cite{Fleischer:1999pa}. According to the analysis
in \cite{Brod:2013sga} the ultimate theoretical error in the determination of $\gamma$ should be far  below experimental sensitivity in the near fruture.
See also \cite{Rout:2019koe,DeBruyn:2022zhw}. As far as $\beta$ is concerned
see detailed discussion in the Belle II book \cite{Belle-II:2018jsg}. The reduction of the error down to the one anticipated in (\ref{CKMoutput26}) should be possible, although here some NP infection could be present. The rapid tests
proposed in \cite{Buras:2022wpw,Buras:2021nns,Buras:2022qip} can help in clarifying it.
\\[1.0cm]  
{\bf Acknowledgements}
I would like to thank Mohamed Zied Jaber for checking
Tables~\ref{tab:1} and \ref{tab:2}.
 Financial support from the Excellence Cluster ORIGINS,
funded by the Deutsche Forschungsgemeinschaft (DFG, German Research
Foundation), 
Excellence Strategy, EXC-2094, 390783311 is acknowledged.

\renewcommand{\refname}{R\lowercase{eferences}}

\addcontentsline{toc}{section}{References}

\bibliographystyle{JHEP}

\small

\bibliography{Bookallrefs}

\end{document}